\documentclass[preprint]{aastex63}

\usepackage{multirow}
\usepackage{amsmath}
\usepackage{appendix}

\shorttitle{Data-constrained Simulation of the Rotation and Eruption of a Flux-Rope System}

\begin{document}

\title{Rotation and Confined Eruption of a Double Flux-Rope System}

\correspondingauthor{Y. Guo}
\email{guoyang@nju.edu.cn}

\author[0000-0002-5690-2244]{X. M. Zhang}
\affiliation{School of Astronomy and Space Science, Nanjing University, Nanjing 210023, China}
\affiliation{Key Laboratory of Modern Astronomy and Astrophysics (Nanjing University), Ministry of Education, Nanjing 210023, China}

\author[0000-0002-4205-5566]{J. H. Guo}
\affiliation{School of Astronomy and Space Science, Nanjing University, Nanjing 210023, China}
\affiliation{Key Laboratory of Modern Astronomy and Astrophysics (Nanjing University), Ministry of Education, Nanjing 210023, China}
\affiliation{Centre for mathematical Plasma Astrophysics, Department of Mathematics, KU Leuven, Celestijnenlaan 200B, B-3001 Leuven, Belgium}

\author[0000-0002-9293-8439]{Y. Guo}
\affiliation{School of Astronomy and Space Science, Nanjing University, Nanjing 210023, China}
\affiliation{Key Laboratory of Modern Astronomy and Astrophysics (Nanjing University), Ministry of Education, Nanjing 210023, China}

\author[0000-0002-4978-4972]{M. D. Ding}
\affiliation{School of Astronomy and Space Science, Nanjing University, Nanjing 210023, China}
\affiliation{Key Laboratory of Modern Astronomy and Astrophysics (Nanjing University), Ministry of Education, Nanjing 210023, China}

\author[0000-0003-3544-2733]{Rony Keppens}
\affiliation{Centre for mathematical Plasma Astrophysics, Department of Mathematics, KU Leuven, Celestijnenlaan 200B, B-3001 Leuven, Belgium}

\begin{abstract}

We perform a data-constrained simulation with the zero-$\beta$ assumption to study the mechanisms of strong rotation and failed eruption of a filament in active region 11474 on 2012 May 5 observed by Solar Dynamics Observatory and Solar Terrestrial Relations Observatory. The initial magnetic field is provided by nonlinear force-free field extrapolation, which is reconstructed by the regularized Biot-Savart laws and magnetofrictional method. Our simulation reproduces most observational features very well, e.g., the filament large-angle rotation of about $130 ^{\circ}$, the confined eruption and the flare ribbons, allowing us to analyze the underlying physical processes behind observations. We discover two flux ropes in the sigmoid system, an upper flux rope (MFR1) and a lower flux rope (MFR2), which correspond to the filament and hot channel in observations, respectively. Both flux ropes undergo confined eruptions. MFR2 grows by tether-cutting reconnection during the eruption. The rotation of MFR1 is related to the shear-field component along the axis. The toroidal field tension force and the non-axisymmetry forces confine the eruption of MFR1. We also suggest that the mutual interaction between MFR1 and MFR2 contributes to the large-angle rotation and the eruption failure. In addition, we calculate the temporal evolution of the twist and writhe of MFR1, which is a hint of probable reversal rotation.

\end{abstract}


\section{Introduction} \label{sec:intro}
Magnetic flux ropes (MFRs) are common configurations in the solar atmosphere. They have many manifestations in the corona, such as filaments/prominences, sigmoids and hot channels \citep{Cheng2017}. Moreover, magnetic flux ropes are generally the cores of coronal mass ejections (CMEs), which are among the most violent eruptions in the solar system. When a CME propagates in the interplanetary space, it is generally termed an interplanetary coronal mass ejection (ICME). About one third of ICMEs, which have a clear flux-rope structure, are magnetic cloud \citep[MC;][]{Burlaga1981}. Furthermore, the southward magnetic field component carried by ICMEs is the major source of geomagnetic storms, which is determined by the orientation of coronal MFRs. It indicates that geomagnetic effects of ICMEs are highly related to the orientations of MFRs and whether they could escape to the interplanetary space. Therefore, to forecast the space weather accurately, it is crucial to figure out the rotation mechanisms of an MFR and whether it could erupt successfully \citep{Shen2022}.

Previous studies indicate that the rotation of a flux rope is highly related to its magnetic helicity and topology \citep{Green2007, Liu2012, Zhou2022}. Many researches reveal that the filament/prominence rotation about its ascending direction depends on its chirality and the helicity sign of the active region \citep{Ouyang2017, Zhou2020, Lynch2009}. For example, a filament in the northern/southern hemisphere generally possesses the dextral/sinistral chirality, corresponding to the negative/positive helicity of the flux rope \citep{Ouyang2017}, and is preferred to rotate counter-clockwise/clockwise \citep{Green2007, Zhou2020}. These observational phenomena can be interpreted by one of the types of ideal MHD instability, namely, helical kink instability (KI), which occurs when the twist of a flux rope is larger than a certain threshold \citep{Kruskal1954}. Due to the conservation of helicity in ideal MHD, the twist of a flux rope might be transferred to its axis writhe, corresponding to the observed rotation of the filament \citep{Torok2010, Zhou2020}. Even so, not all filaments satisfying KI can erupt successfully, since another instability, namely, torus instability (TI), also plays a crucial role in determining the fate of an eruption \citep{Kliem&Torok2006}. A flux rope exceeding the threshold of KI might be confined in the lower corona owing to unsatisfied TI, which is the failed kink regime \citep{Torok&Kliem2005}. Furthermore, the change in the toroidal field tension force produced by the interaction between the guide field and flux-rope poloidal current could also constrain the eruption, which is termed the failed torus \citep{Myers2015}. Besides, a non-axisymmetrical force caused by the radial magnetic field also can constrain the eruption, which is another essential force to consider \citep{Zhong2021}.

Albeit a single flux rope is already able to explain many aspects of the eruption phenomena,
both observations and data-based extrapolations imply that there might exist more than one flux rope in an active region. For instance, \citet{Kumar2010} observed that two adjacent filaments merge in their middle portions at first, and then bounce reversely in perpendicular directions, which can be explained by the reconnection of two flux ropes. \citet{Liu2012} found a pair of separated filaments above one polarity inversion line (PIL), namely, a double-decker filament. \citet{Liut2018} studied a sigmoid system comprised of a filament and hot channel with magnetic field extrapolations and found that the eruption of the overlying flux rope (hot channel) can suppress the rising of the low-lying flux rope (filament). Furthermore, a filament can also be triggered to ascend by tether-cutting reconnection of two adjacent sheared arcades underneath \citep{Zou2019}. In addition to observations, some theoretical instability regimes depend on interactions between multiple current channels \citep{Keppens2019}. \citet{Linton2001} summarized four scenarios of interactions between two flux ropes (bounce, merging, slingshot and tunnel) by performing a parameter survey. \citet{Keppens2014} carried out a three-dimensional (3D) MHD simulation of two adjacent antiparallel current channels, and found that two flux ropes can repel each other with rotational motions, reflecting the tilt-kink instability. Two flux ropes can also interact indirectly. \citet{Torok2011} found that the eruption of a flux rope under a pseudo-streamer structure is liable to be initiated by the eruption of a flux rope nearby, which is termed sympathetic eruption. However, the above studies have limitations. On the one hand, the majority of the present models focus on the interaction between flux ropes formed before the eruption, but some flux ropes are formed during eruption due to magnetic reconnection \citep{Wang2017}. On the other hand, theoretical simulations are based on ideally parameterized magnetic fields. However, the actual magnetic fields topologies are more intricate, so it is essential to perform data-based numerical simulations to study these issues.

Here, to study the rotation of a flux rope, we perform a data-constrained MHD simulation for the filament eruption occurring on 2012 May 5, which has a relatively large rotation angle during the eruption. The initial magnetic field is obtained with nonlinear force-free field (NLFFF) extrapolation, and the bottom boundary is provided by the observed vector magnetogram. We focus on two goals in this paper. First, we expect to reproduce as many typical features in observations as possible, which are used to assess whether the simulation is reasonable. Second, we attempt to reveal the underlying physical mechanisms behind the observations, such as the role of the magnetic topology evolution and magnetic reconnection. The observational features and numerical setup are described in Section \ref{sec:obsn}. The simulation results are presented in Section \ref{sec:res}, which is followed by the summary and discussions in Section \ref{sec:sum}.

\section{Observations and Numerical Setup} \label{sec:obsn}

The event we study is a filament eruption accompanied by a weak and ungraded flare, which starts at 17:35 UT and peaks at 17:41 UT on 2012 May 5, and is hosted in NOAA Active Region 11474 (more information of the flare can be found in the website "Solar Flares Query Page"\footnote{https://data.nas.nasa.gov/helio/portals/solarflares/webapp.html}). The filament material presents distinct radiation features, reflecting the morphology evolution of the flux rope during the eruption. Moreover, this event is captured by the Atmospheric Imaging Assembly \citep[AIA;][]{Lemen2012} on board Solar Dynamics Observatory (\emph{SDO}) and the Extreme Ultraviolet Imager \citep[EUVI;][]{Wuelser2004} on board Solar Terrestrial Relations Observatory (\emph{STEREO}) simultaneously, so we can observe it with different viewing angles. Considering the analysis of observations and magnetic structure \citep{Zhou2019,Guojh2021a}, this eruption shows many unusual characteristics. First, the filament rotates counter-clockwise for a relatively large angle ($130 ^{\circ}$) while it rises. Second, the filament fails to erupt. Here, we focus on the mechanisms causing the rotation and failed eruption of this event. To investigate these problems, it is essential to reproduce the observations and analyze the magnetic topology evolution with a data-based simulation.

Similar to previous works \citep{Guo2019, Zhong2021,Guoy2021}, we adopt the zero-$\beta$ MHD model, which is described by the governing equations as follows:
\begin{eqnarray}
 && \frac{\partial \rho}{\partial t} +\nabla \cdot(\rho \boldsymbol{v})=0,\label{eq1}\\
 && \frac{\partial (\rho \boldsymbol{v})}{\partial t}+\nabla \cdot(\rho \boldsymbol{vv}-\boldsymbol{BB})+\nabla (\frac{\boldsymbol{B}^2}{2})=0,\label{eq2}\\
 && \frac{\partial \boldsymbol{B}}{\partial t} + \nabla \cdot(\boldsymbol{vB-Bv})=0,\label{eq3}
\end{eqnarray}
where ${\rho}$ represents the density, $\boldsymbol{v}$ is the velocity, and $\boldsymbol{B}$ is the magnetic field. All of these physical quantities are normalized by their typical coronal values as follows: $L_{0}=1.0\times 10^{9}$ cm, $B_{0}=2.0$ G, $t_{0}=85.9$ s, $\rho_{0}=2.3\times 10^{-15}$ $\rm g\ cm^{-3}$, $v_{0}=1.2\times 10^{7}$ $\rm cm^{-1}$. In this zero-$\beta$ model, the gravity, gas pressure and energy equations are omitted so that the dynamics of the flux rope are fully driven by the Lorentz force next to inertial effects. In particular, the explicit resistivity is set to zero since reconnection can still occur due to inherent numerical diffusion. 

\begin{figure}[ht!]
\centering
\includegraphics[scale=0.35]{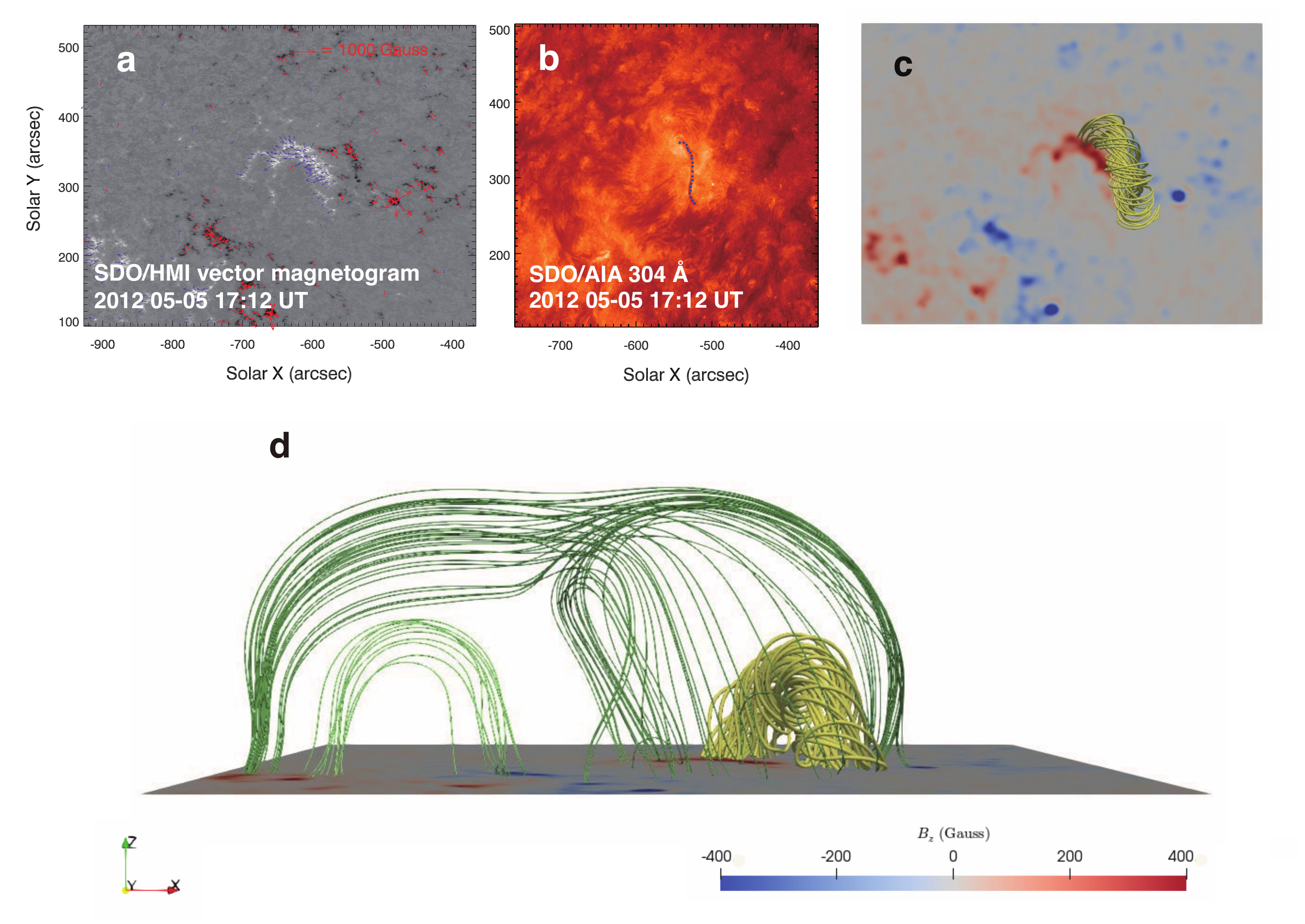}
\caption{(a) \emph{SDO}/HMI remapped vector magnetogram in the local Cartesian coordinate system at 17:12 UT on 2012 May 5, where the purple (red) arrows represent the directions of the local horizontal field in positive (negative) polarities. (b) \emph{SDO}/AIA 304 {\AA} image at the same time as (a). Dark blue dots outline the filament, which is taken as the path of the inserted flux rope. (c) Top view of the flux rope (yellow lines), which is constructed by RBSL and MF methods. (d) Side view of the flux rope and background field (green lines).}
\label{fig1}
\end{figure}

The initial coronal magnetic field is obtained by NLFFF extrapolation, which makes use of the regularized Biot-Savart laws \citep[RBSL;][]{Titov2018}. The bottom vector magnetic field uses observations at 17:12 UT on 2012 May 5 provided by the Helioseismic and Magnetic Imager \citep[HMI;][]{Scherrer2012, Schou2012} on board \emph{SDO}, which is processed to correct for projection effect \citep{Guo2017b} and the removal of the net Lorentz force and torque \citep{Wiegelmann06}. The remapped vector magnetic field is shown in Figure \ref{fig1}a. To reconstruct the coronal magnetic field containing a flux rope, we adopt the RBSL method proposed by \citet{Titov2018} and implemented by \citet{Guo2019b}. The RBSL model is controlled by four parameters, i.e., the flux-rope path, $\mathcal{C}$, minor radius, $a$, toroidal flux, $F$, and electric current, $I$. Among them, the minor radius is set as the observed width of the filament, approximately 7 Mm. Regarding the flux-rope path, similar to \citet{Guojh2021a}, we outline the projection path of the filament in the AIA 304 {\AA} image at first, as shown by dark blue circles in Figure \ref{fig1}b. Then, we adopt the triangulation method to measure the apex height of the filament ($h$), and the heights of rest points are fitted by an arc. It is noted that, since filament material is expected to be deposited at the bottom of MFR \citep{Guo2022}, the apex height of the flux-rope axis is set as the sum of the apex height of the filament and the minor radius. Once the flux-rope path and the minor radius are determined, the toroidal flux $F_{0}$ can be estimated by the mean value of the unsigned magnetic fluxes at the two conjugate footpoints, that is, $F_0= (|F_+|+|F_-|)/2=1.5\times 10^{20}$ Mx. In this event, we take $F=4.5F_{0}$ after many numerical tests, which can reproduce the majority of observational features relatively well. Then the electric current $I$ is derived from the equilibrium condition, i.e., Equation (12) in \citet{Titov2018}. Particularly, we take a negative sign for current $I$ since the filament is dextral, according to the draining sites and hemisphere rule \citep{Chen2014, Ouyang2017}. Finally, we close the flux-rope path by adding a mirror sub-photosphere arc, which is used to keep the normal magnetic field on the photosphere unchanged. After that, we embed the above flux rope into the potential magnetic field extrapolated with the Green's function method \citep{Chiu1977} and relax it to a force-free state with the magnetofrictional (MF) method \citep{Guo2016a, Guo2016b}. After relaxation, the force-free metric is $\sigma_{J}=0.241$ and the divergence-free metric is $\langle |f_{i}|\rangle = 7.97 \times 10^{-4}$, which are acceptable values for the majority of published data-constrained model extrapolations \citep{Guo2016a, Guo2016b}. Figures \ref{fig1}c and \ref{fig1}d show the top and side views of the coronal magnetic field, respectively. The flux-rope minor radius expands strongly after MF and its morphology resembles the observed filament. The flux rope is located under a fan-spine structure. So far, we obtained the initial magnetic field of the time-dependent, 3D simulation.

There are some things to stress about the choice of the initial state. First, the force-free state after MF is an approximate equilibrium state but not a stable state. This means that the flux rope may contain unstable flux and start to evolve under small or large (for nonlinear stability) disturbances. Considering the complexity and transverse field errors of observed photosphere magnetic field, the relaxed force-free magnetic fields cannot reach a perfect equilibrium state. The criterion for choosing a proper initial state includes minimizing the force-free metric and the divergence-free metric and keeping the bottom magnetic field as observed. Although the force-free metric and divergence-free metric cannot reach exactly zero, both of them are in a reasonable range after a sufficient relaxation. Here, we use the state after 40000-step relaxation as the initial condition of the simulation.

The initial density distribution is given by stratified atmospheric model, which is the same as \citet{Guo2019}:
\begin{eqnarray}
&& T=\left\{
\begin{array}{lccl}
T_0               & &       &{0 \le h<h_0}\\
k_{_{T}}(h-h_0)+T_0    & &       &{h_0 \le h<h_1}\\
T_1               & &       &{h_1 \le h<z_{max}}
\end{array}\right., \label{eq4}\\
&& {\qquad \quad} g(R_\sun+h)^2=g_0 R_\sun^2 ,\label{eq5}\\
&& {\qquad \quad} \frac{\text{d}(\rho T)}{\text{d}h} = -\rho {g ,\label{eq6}}
\end{eqnarray} 
where ${T_{0}= 0.006}$, ${T_{1}= 1}$, ${h_{0}=0.35}$, ${h_{1}=1.0}$, ${k_{_{T}}=(T_{1}-T_{0})/(h_{1}-h_{0})}$, ${R_\sun=69.55}$ (solar radius), ${g_{0}=0.2}$ in dimensionless units. We adopt a temperature-height distribution and calculate the gravity acceleration distribution. With the assumption of the hydrostatic atmosphere, the density profile can be derived with given density ($\rho_{0}= 1.0 \times 10^{8}$) at the bottom. We note that the only effect of such temperature and gravity distribution is to derive the initial density distribution in our simulation and there is no temperature and gravity in the zero-$\beta$ model, which are zero. The initial velocity is also set to be zero everywhere.

The above equations in the local Cartesian coordinate system are numerically solved with the Message Passing Interface Adaptive Mesh Refinement Versatile Advection Code \citep[MPI-AMRVAC\footnote{http://amrvac.org},][]{Xia2018, Keppens2020,AMRVAC2023}. The computational domain is ${[x_{min},x_{max}]\times[y_{min},y_{max}]\times[z_{min},z_{max}]=[-205,205]\times[-154,154]\times[1,308]}$ Mm, which is resolved by  ${280\times210\times210}$ cells, i.e., we use a uniform fixed grid here (no AMR). The HLL Riemann solver, a three-step time integration and Koren limiter are adopted and two ghostcells are used to set the boundary conditions. Similar to \citet{Guoy2021}, we adopt the data-constrained bottom boundary, where the magnetic field in the inner ghost cell is provided by the observed vector magnetic field, while the outer layer is calculated by a second-order zero-gradient extrapolation, and the velocity is set to zero in these layers. Hence, the bottom boundary is not driven by successive magnetograms. Regarding the other five boundaries, the magnetic field is computed by a second-order zero-gradient extrapolation, the density is set as the initial condition, and the velocity is set as zero there.

\section{Result} \label{sec:res}
\subsection{Comparisons between Simulation and Observations} \label{subsec:compar}

To compare the simulated flux rope with observations, we perform a back-projection for the vector magnetic field and the on-disk location, so that they could be viewed either in \emph{SDO} or \emph{STEREO-B} perspectives. The rotation matrix of the back-projection is described by the matrix multiplication of elementary rotations, i.e., ${R_x(-B_0)R_y(-L)R_x(B_1)}$ \citep[refer to][for more details]{Guo2017b}. Among them, ${B_0}$ is the latitude of the disk center, ${L}$ and ${B_1}$ are the longitude and latitude of the projected center, respectively. For the event in this study, ${(B_0, L, B_1)}$ for \emph{SDO}/AIA and \emph{STEREO-B}/EUVI on 2012 May 5 are ${(3.7^{\circ}, -35.5^{\circ}, 16.0^{\circ})}$ and ${(-3.6^{\circ}, 83.0^{\circ}, 16.0^{\circ})}$, respectively.

\begin{figure}[ht!]
\centering
\includegraphics[scale=0.4]{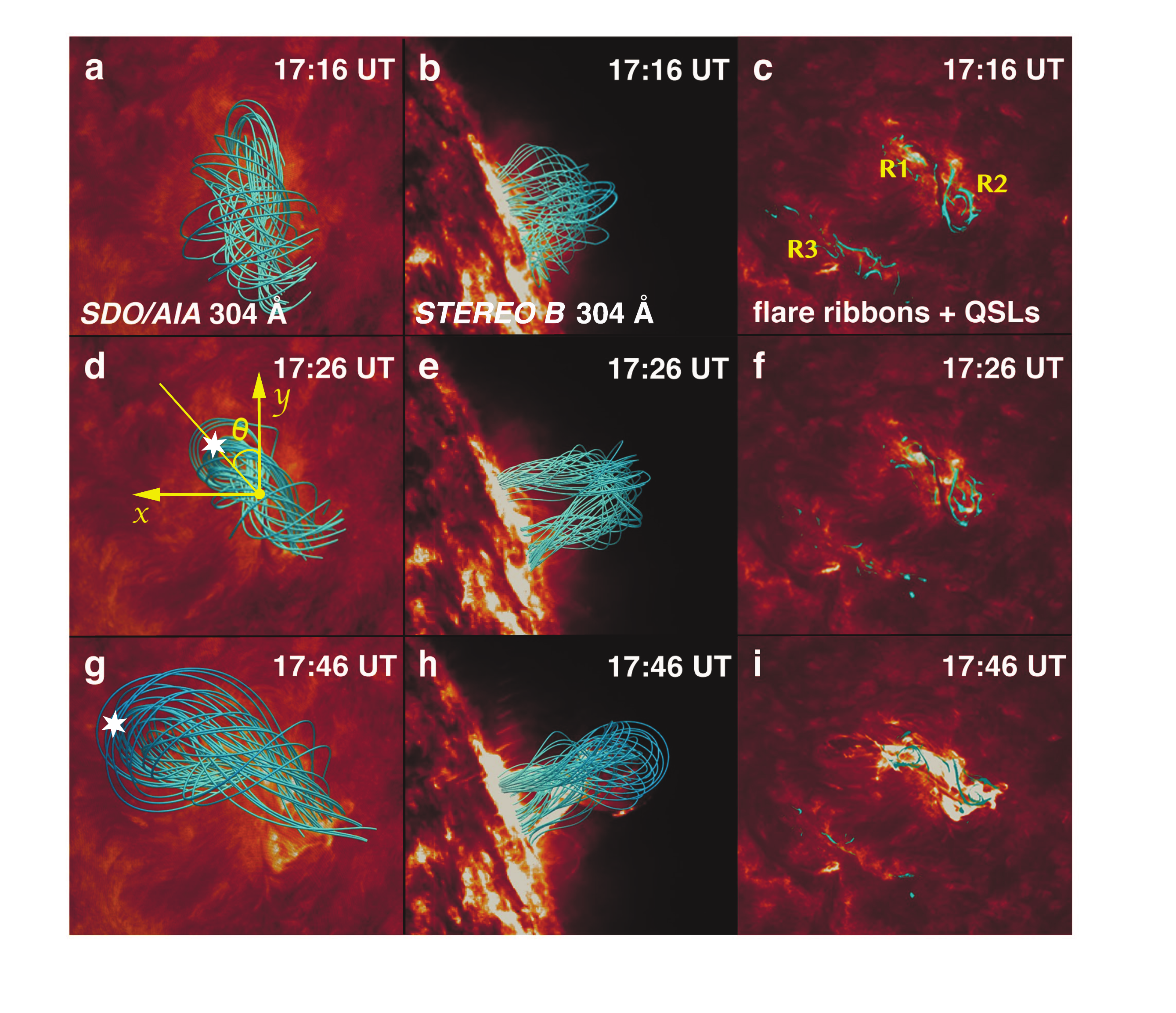}
\caption{Comparisons between simulation results and observations at 304 {\AA}, where the cyan lines represent the flux rope. The three rows from the top to bottom display the observations and simulations at 17:16, 17:26, and 17:46 UT on 2012 May 5. The left (a, d, g) and middle (b, e, h) columns show the simulated flux rope overlaid on 304 {\AA} images viewed from \emph{SDO} and \emph{STEREO-B}, respectively. The right column (c, f, i) illustrates the photosphere QSLs (lg$Q>$2) and flare ribbons on \emph{SDO}/AIA 304 {\AA} images. An animation is available showing the above evolutions both without (the first row in the animation) and with (the second row) the flux ropes superimposed from 17:12 UT to 17:50 UT. }
\label{fig2}
\end{figure}

Figure \ref{fig2} shows the evolution of the flux rope after the back-projection, which is overlaid on 304 {\AA} images with the viewing angles from \emph{SDO} (Figures \ref{fig2}a, \ref{fig2}d and \ref{fig2}g) and \emph{STEREO-B} (Figures \ref{fig2}b, \ref{fig2}e and \ref{fig2}h). Snapshots in the top and side views reproduce many observational features, in particular, the eruption direction, rotation and morphology of the flux rope. For example, in top views, both the simulated flux rope and the observed filament show a counter-clockwise rotation, which is in accord with the relationship between the rotation and chirality \citep{Green2007}. In side views, the inverse-$\gamma$ morphology in observations is reproduced by our simulation fairly well.

Then we compute the squashing factor ($Q$) based on the method proposed by \citet{Scott2017}, which is realized by an open-source routine (K-QSL) provided by Kai E. Yang \footnote{https://github.com/Kai-E-Yang/QSL}. Quasi-separatrix layers (QSLs) are places with drastic magnetic field linkage changes, which are layers of high $Q$ ($Q\gg2$). QSLs depict favorite regions where reconnection could occur \citep{Priest1995, Titov2002}, which are usually cospatial with flare ribbons \citep{Guo2019, Zhong2021, Guojh2023}. Thus, the spatial relationship between QSLs and flare ribbons is a reliable validation of the simulation. As shown in Figure \ref{fig2}c, simulated QSLs reproduce successfully three major ribbons, including two parallel ribbons (R1 and R2) and a remote ribbon (R3). Additionally, the footpoints of the flux rope are almost located at the QSL hook and the concave side of the ribbon \citep{Aulanier2019}. The spatial coherency between QSLs and flare ribbons strongly indicates that the topology evolution of our simulation captures the main aspects of this event.

\begin{figure}[ht!]
\centering
\includegraphics[scale=0.65]{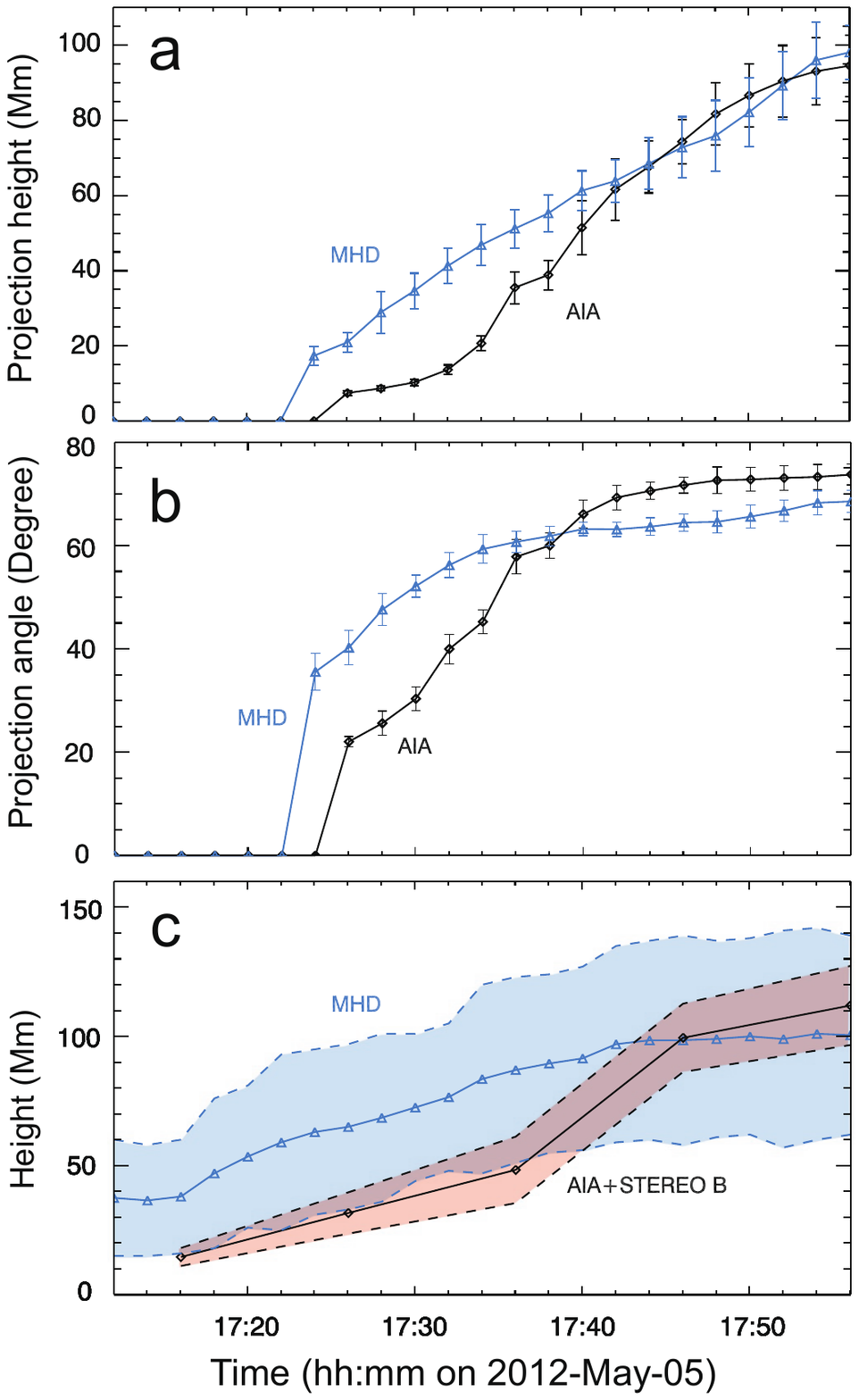}
\caption{Quantitative comparisons between the simulated flux rope and the erupted filament in observations, where the blue lines with triangle signs represent the simulated flux rope, and the black lines with diamond signs represent the filament in observations. Panels (a) and (b) display time-distance profiles and time-angle profiles, respectively. The distances and angles of the simulated flux rope and the observed filament are measured in the viewing angle from \emph{SDO}. (c) Time-height profiles in 3D of the simulated flux rope and the filament, where the blue (red) shadings represent the flux-rope (filament) boundary. The error bars shown in (a) and (b) are calculated by the uncertainty with repeated measurements of ten times.}
\label{fig3}
\end{figure}

Next, we compare quantitatively the kinematics of the simulated flux rope and the filament in observations. First, we compare the projection height viewed from \emph{SDO} (Figure \ref{fig3}a), which is measured by the distance between the flux rope/filament apex and a reference point. We repeat the measurements ten times to reduce the errors. Then, we compare the projection angles between the simulation and observation, as shown in Figure \ref{fig3}b. The projection angle is measured by the angle between the $y$-axis and the line connecting the reference point and flux rope/filament apex, which is illustrated in Figure \ref{fig2}d. This angle indicates the deflection between the eruption direction and the $y$-axis but not a 3D rotation angle about its rising direction. Combining the projection height (Figure \ref{fig3}a) and the projection angle (Figure \ref{fig3}b), we can verify if the ejection direction in our simulation is reasonable or not. It is shown that the simulated flux rope coincides well with observations, in particular the ultimate height and angle when the flux rope stops rising and rotating. The only difference is that the simulated flux rope starts to rise about 2 minutes earlier than the observed filament. Additionally, we measure the 3D height in the observation with the triangulation method and that of the simulated flux rope in the 3D simulation domain (Figure \ref{fig3}c). We find that the simulated flux rope is approximately 25 Mm higher than the filament, consistent with the expectation that the filament is located at the bottom of the flux rope \citep{Guo2022}. 

\begin{figure}[ht!]
\centering
\includegraphics[scale=0.55]{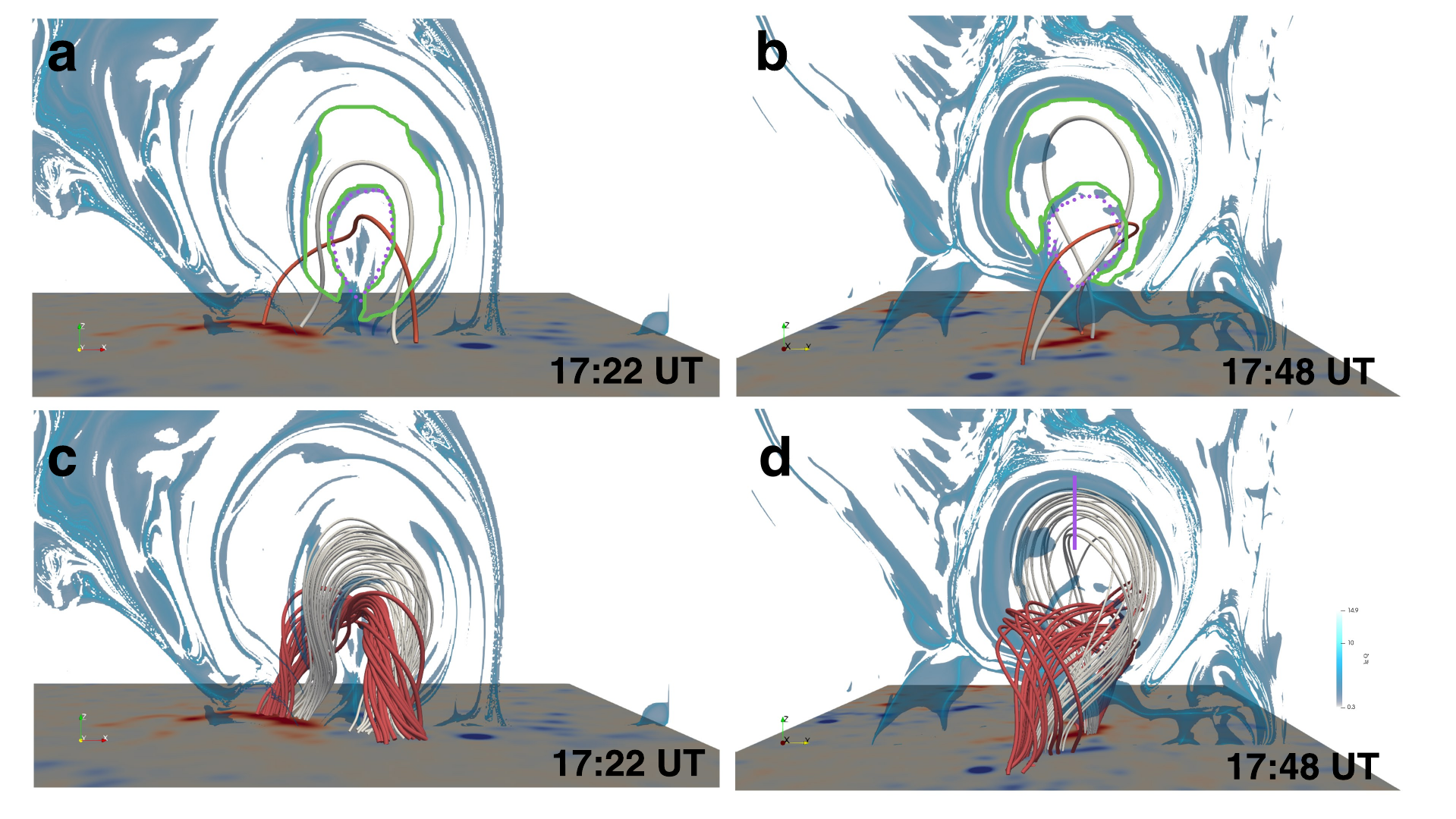}
\caption{QSLs on a slice and the two flux rope configuration. Panels (a) and (b) show the QSLs at 17:22 UT and 17:48 UT, respectively. The regions encircled by the green solid lines (purple dotted lines) indicate the cross section of the gray (red) flux rope in panels (c) and (d). The gray and red field lines in panels (a) and (b) mark the axis locations of MFR1 and MFR2, respectively. Panels (c) and (d) show the configuration of two flux ropes on the background of QSLs at the same time as (a) and (b), respectively. The purple vertical line in panel (d) labels the diagnostic line for the force decomposition in Fig. \ref{fig7}. The directions of QSL cross sections are different in different panels, which is perpendicular to $y$-axis in panels (a) and (c) and is perpendicular to $x$-axis in panels (b) and (d). The total height extent of the Q maps is 138 Mm in panels (a) and (c) and 198 Mm in panels (b) and (d).}
\label{fig4}
\end{figure}

\subsection{Double flux rope system formation and topology evolution} \label{subsec:mtf}

During the simulation, another flux rope forms beyond the minor radius of the inserted flux rope. Considering different topologies and evolution processes, the field lines are classified into two flux ropes. Figures \ref{fig4}a and \ref{fig4}b show the QSLs on a vertical diagnostic plane parallel to the $xz$ and $yz$ planes, which contain at least two separated areas in the whole flux rope region. The boundaries of two flux ropes are outlined in Figures \ref{fig4}a and \ref{fig4}b in different colors and line styles. Then, we trace magnetic field lines at the centers of two main separated areas and find two flux ropes. In Figures \ref{fig4}c and \ref{fig4}d, the flux-rope system and the QSL distribution are overlapped and the different orientations of the two flux ropes (in gray and red) can be displayed clearly.

\begin{figure}[ht!]
\centering
\includegraphics[scale=0.25]{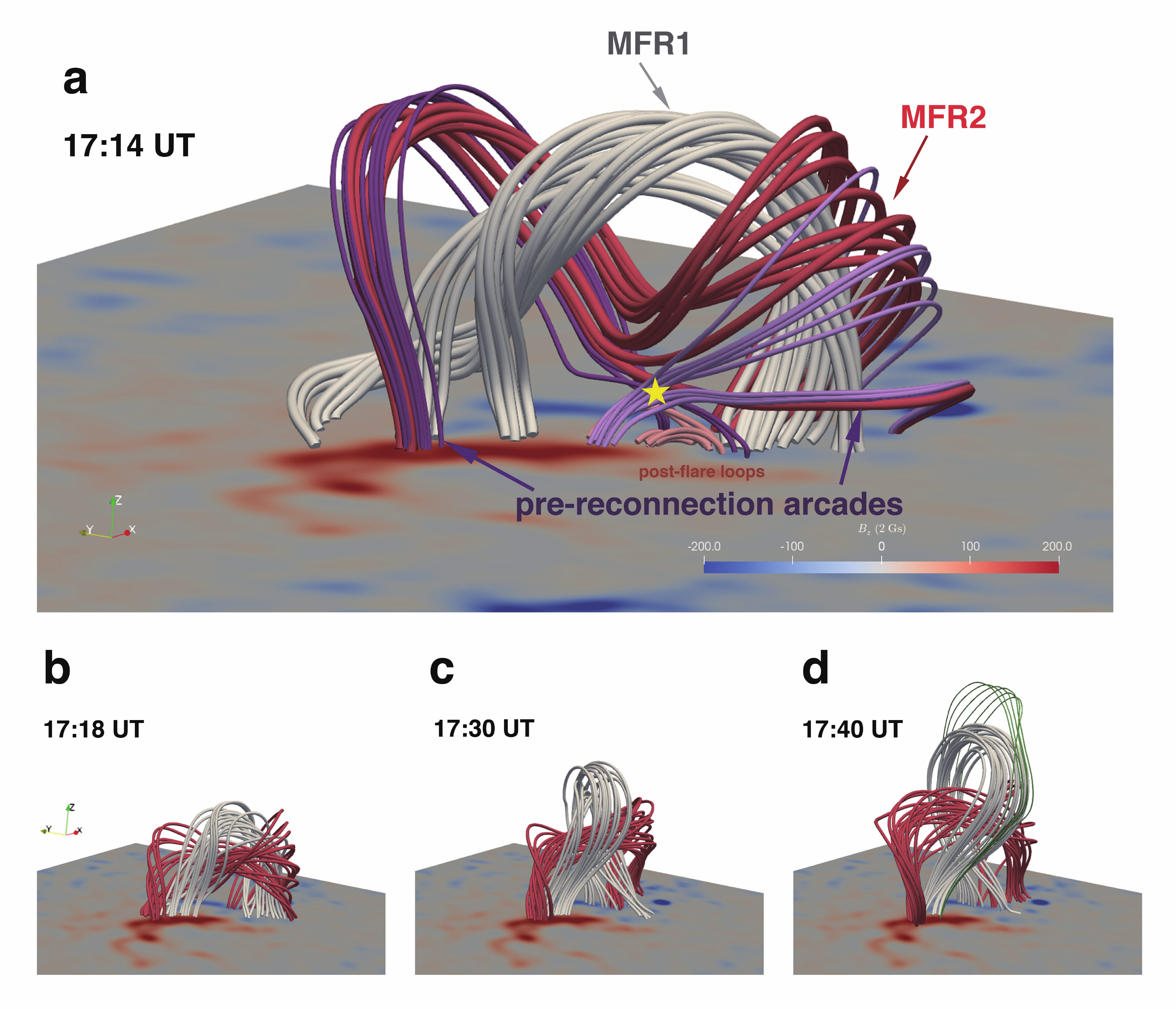}
\caption{Evolution of the flux-rope system, where gray lines represent the inserted flux rope (MFR1), and red lines represent the flux rope formed during the eruption (MFR2). (a) Illustration of the tether-cutting reconnection at 17:14 UT on 2012 May 5. The yellow pentagram represents the reconnection site, purple field lines represent the pre-reconnection arcades and the pink arcades underneath represent the post-flare loops. (b--d) Snapshots of the evolution of MFR1 and MFR2 at 17:18, 17:30 and 17:40 UT on 2012 May 5. The thin green lines in (d) represent some background overlying field lines.}
\label{fig5}
\end{figure}

Figure \ref{fig5}a shows some sample field lines at 17:14 UT, from which we exhibit two flux ropes and highly sheared arcades. The flux rope colored in gray (MFR1) corresponds to the observed filament, which is constructed by the RBSL method. The flux rope colored in red (MFR2) grows up during the eruption, which originates from a pre-existing hyperbolic flux tube (HFT) below the body of MFR1 and is formed gradually by the injection of flux caused by the tether-cutting reconnection \citep{Moore1980} of two bunches of highly sheared arcades underneath it. We trace the twisted field lines at the apex of the current sheet above the HFT and identify them as MFR2. Then, we show the evolution of the two flux ropes, as illustrated in Figures \ref{fig5}b--\ref{fig5}d. MFR1 exhibits an evident counter-clockwise rotation during the eruption and presents an inverse-$\gamma$ morphology at 17:40 UT. The orientation of MFR1's apex at that time is almost parallel to that of the overlying background field lines. The average rise velocity of MFR1 is about 10 $\rm km \  s^{-1}$ from 17:46 UT to 17:56 UT, in comparison with the maximum rise velocity of 65 km $\rm km \  s^{-1}$ at 17:18 UT, indicating that it is a failed eruption. MFR2 expands in major radius with only little rotation and its final rise velocity is about $-9$ $\rm km \  s^{-1}$, indicating that MFR2 is also confined. 

In general, magnetic helicity is invoked to describe the complexity of magnetic fields, which can be divided in contribution from writhe ($W_r$) and twist ($T_w$) for a single flux rope \citep{Valori2016}. Among them, the writhe denotes the deformation of the flux-rope axis, and the twist describes how field lines wind around an axis. To understand MFR1's rotation, we investigate the evolution of these two parameters. We take the field line that passes through the spot of maximum current at each time step, which means the axis may vary as time progresses and $T_w$ and $W_r$ are approximations calculated in this way. We calculate its polar writhe with the following steps \citep{Berger&Prior2006}. First, as the open curve may have several extrema in height, we split it in monotonically height-varying pieces $i$ by several local extrema in $z$ direction, where $[z_i^{min}, z_i^{max}]$ represents $z$-range of piece $i$. Then, we split the polar writhe into two parts, namely, the local writhe ($W_{local}$) and the nonlocal writhe ($W_{nonlocal}$), which can be calculated with the following formulae \citep{Berger&Prior2006},
\begin{eqnarray}
&& {\boldsymbol{W}_{local}=\frac{1}{2\pi}\sum_{i=1}^n \int_{z_i^{min}}^{z_i^{max}} \frac{1}{1+\vert \boldsymbol{T}_i\cdot \boldsymbol{z} \vert} (\boldsymbol{T}_i \times \frac{\text{d} \boldsymbol{T}_i}{\text{d} z})_{z} \text{d}z },\label{eq7}\\
&& {\boldsymbol{W}_{nonlocal}=\sum_{i=1}^{n}\sum_{j=1,j\neq i}^{n} \frac{\sigma_{ij}}{2\pi} \int^{min(z_i^{max},z_j^{max})}_{max(z_i^{min},z_j^{min})} \frac{\text{d} \Theta_{ij}}{\text{d}z} \text{d}z },\label{eq8}
\end{eqnarray}
where $n$ is the total number of pieces, $\boldsymbol{T}_i$ is the tangent vector at a certain $z$ on the curve piece, $\boldsymbol{z}$ is the unit vector in $z$ direction, $\sigma_{ij}$ represents the sign of the terms. If piece $i$ and piece $j$ have the same tending in $z$ direction, $\sigma_{ij}$ is positive; otherwise, $\sigma_{ij}$ is negative. $\Theta_{ij}$ is the angle between $x$-axis (an arbitrary direction perpendicular to $\boldsymbol{z}$) and the vector from $i$ piece to $j$ piece. The total polar writhe $W_r$ is the sum of $W_{local}$ and $W_{nonlocal}$. We note that \citet{Zhou2022} only considers the nonlocal part.

Regarding the twist number ($T_w$), we calculate it with methods similar to previous works \citep{Guo2017b, Guojh2021a}. According to the connectivity variation of the field line and the distribution of the current, we select 250 sample field lines within MFR1, which is wrapped by closed QSLs, although their start points are different at each time step. We calculate their mean twist with the following computational formula \citep{Berger&Prior2006},
\begin{eqnarray}
&& \frac{\text{d}T_w}{\text{d}s} = \frac{1}{2\pi} \boldsymbol{T}\cdot \boldsymbol{V} \times \frac{\text{d}\boldsymbol{V}}{\text{d}s},\label{eq9}
\end{eqnarray}
where $s$ is the arc length between a certain point (A) on the axis of the flux rope and the reference point, $\boldsymbol{T}$ is the unit tangent vector at point A on the axis, $\boldsymbol{V}$ is the unit vector perpendicular to $\boldsymbol{T}$ and pointing to the field line around the axis. The temporal evolution of $W_r$ and $T_w$ is shown in Figure \ref{fig6}, which can be divided into two stages. First,  $|W_r|$ increases and $|T_w|$ decreases, indicating the transfer from the twist to writhe. This stage corresponds to the counter-clockwise rotation in observations. Then, more intriguingly, an inverse transfer occurs in the second stage, that is, $|W_r|$ decreases and $|T_w|$ increases.

To analyze the exact reason for the confined eruption of MFR1, we use our simulated magnetic field at 17:48 UT to calculate the Lorentz force and its components around the axis of MFR1. First, we use the simulated magnetic field and current to calculate the total Lorentz force in the $z$-direction. The sample points for force calculation are located on a vertical line through the apex of MFR1's axis. Then, the vertical Lorentz force is decomposed into different components, based on the field components in the toroidal, poloidal and radial directions and current density components induced by these three field components. The toroidal direction is along the horizontal field direction at the vertical line through the apex point of the MFR1 axis. The poloidal direction is defined by the right-hand screw rule referred to the toroidal direction. The radial direction radiates outward from the MFR1 (axis) to the edge. These three directions are orthogonal to each other, which form the three basis vectors in cylindrical coordinates. Namely, the toroidal, polodial and radial directions correspond to the $z$, $\theta$ and $r$ directions in cylindrical coordinates, respectively. The method to decompose the Lorentz force and the defination of the non-axisymmetry force $F_{N1}$ and $F_{N2}$ are explained in \citet{Zhong2021}. Figure \ref{fig7} shows the force distribution, in which the value is normalized by $f_0=10^{-10} \ \rm dyn\ \rm cm^{-3}$. The toroidal field tension force, $F_\mathrm{T} = \mathbf{e}_\mathrm{z} \cdot (\mathbf{J}_\mathrm{RP} \times \mathbf{B}_\mathrm{T}) $, where $\mathbf{J}_\mathrm{RP} = \nabla \times \mathbf{B}_\mathrm{T}/\mu_0$, and $\mathbf{B}_\mathrm{T}$ represents the total toroidal magnetic field, is downward below the axis. However, the strapping force and the hoop force are upward. The toroidal field tension force changes its direction to upwards across the axis and the other two forces are still upward, which means some other components are needed to confine the eruption. For this purpose, we calculate the non-axisymmetry forces which termed as $F_{N1}$ and $F_{N2}$ in Figure \ref{fig7} and they are also downward. Especially, $F_{N2}$ is downwards all the time with a relatively large value. Finally, to verify the validity of our decomposition, the residual force is used. It is calculated by substracting all the component forces from the total Lorentz force in the $z$-direction. The residual force identically equals to zero everywhere, which indicates the validity our result. 


\section{Summary and Discussion} \label{sec:sum}

We conduct a data-constrained simulation to analyze the filament eruption observed by \emph{SDO} in NOAA AR 11474, on 2012 May 5. Our simulation exploits the open-source code MPI-AMRVAC to solve the zero-$\beta$ MHD equations. As for the numerical setup, the bottom boundary is provided by the vector magnetogram recorded by \emph{SDO}/HMI at 17:12 UT and the initial magnetic field adopts the NLFFF model constructed by RBSL and MF methods. Our simulation results reproduce the dynamic evolution of the observed filament and flare ribbons very well. First, both the morphology and the erupted direction of the flux rope are similar to the observed filament. Moreover, we compare the height and projected angle of the erupting flux rope in the simulation to those in the observation quantitatively, both of which show resemblances in ejection direction and velocity. Second, the QSLs in the simulation represent a good spatial correspondence with the observed flare ribbons, which indicates the reliability of the magnetic topology evolution of our simulation. After that, we analyze the magnetic configuration during the eruption and discover a double-MFR eruption core. The upper flux rope (MFR1) exists before the eruption, which corresponds to the observed filament. The lower flux rope (MFR2) grows gradually by tether-cutting reconnection between two adjacent sheared magnetic arcades during the eruption. Unlike \citet{Liu2016} and \citet{Awasthi2018}, whose models have a configuration with several parallel flux ropes, our simulation has two flux ropes whose orientations differ by a relatively large angle around $150^{\circ}$ during the eruption. It is impossible to select a single plane that can clearly exhibit two axes. In Figure \ref{fig4}, we plot QSLs on a slice which is almost perpendicular to MFR2's axis. It indicates that MFR1 and MFR2 are located in different flux systems. The newly formed flux rope (MFR2) contributes to the rotation and confined eruption of the entire magnetic system. We calculated the evolution of magnetic helicity parameters of MFR1, which shows an obvious conversion between twist and writhe. It describes the rotation quantitatively with the helicity parameters of the flux rope.

\begin{figure}[ht!]
\centering
\includegraphics[scale=0.55]{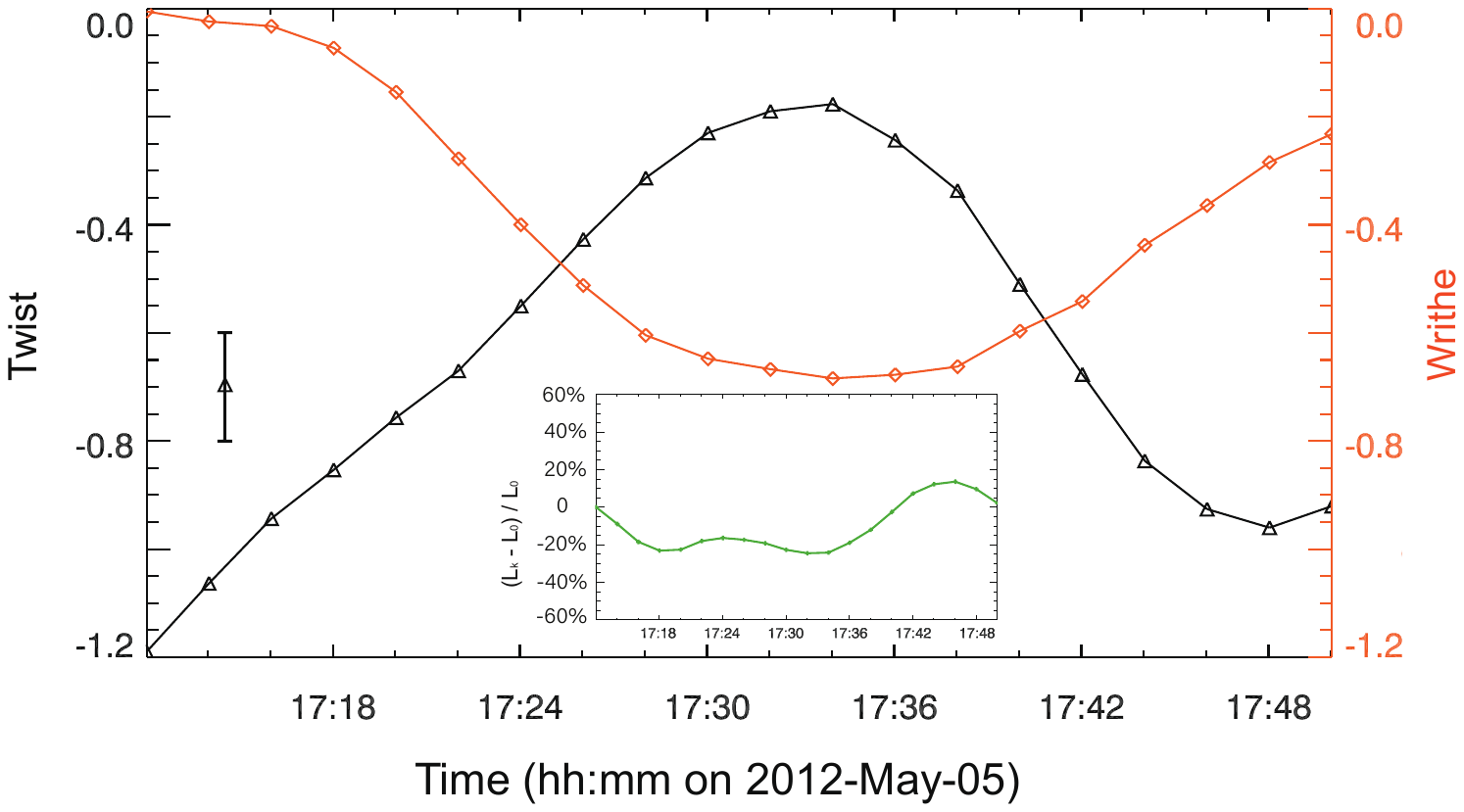}
\caption{Temporal evolution of twist (black line with triangles) and writhe (orange line with diamonds) of MFR1 during the eruption. The error bar refers to the mean value of the standard deviation of selected field lines for all data points. The inside panel shows the changing rate of the linking number during the eruption.}
\label{fig6}
\end{figure}

\begin{figure}[ht!]
\centering
\includegraphics[scale=0.7]{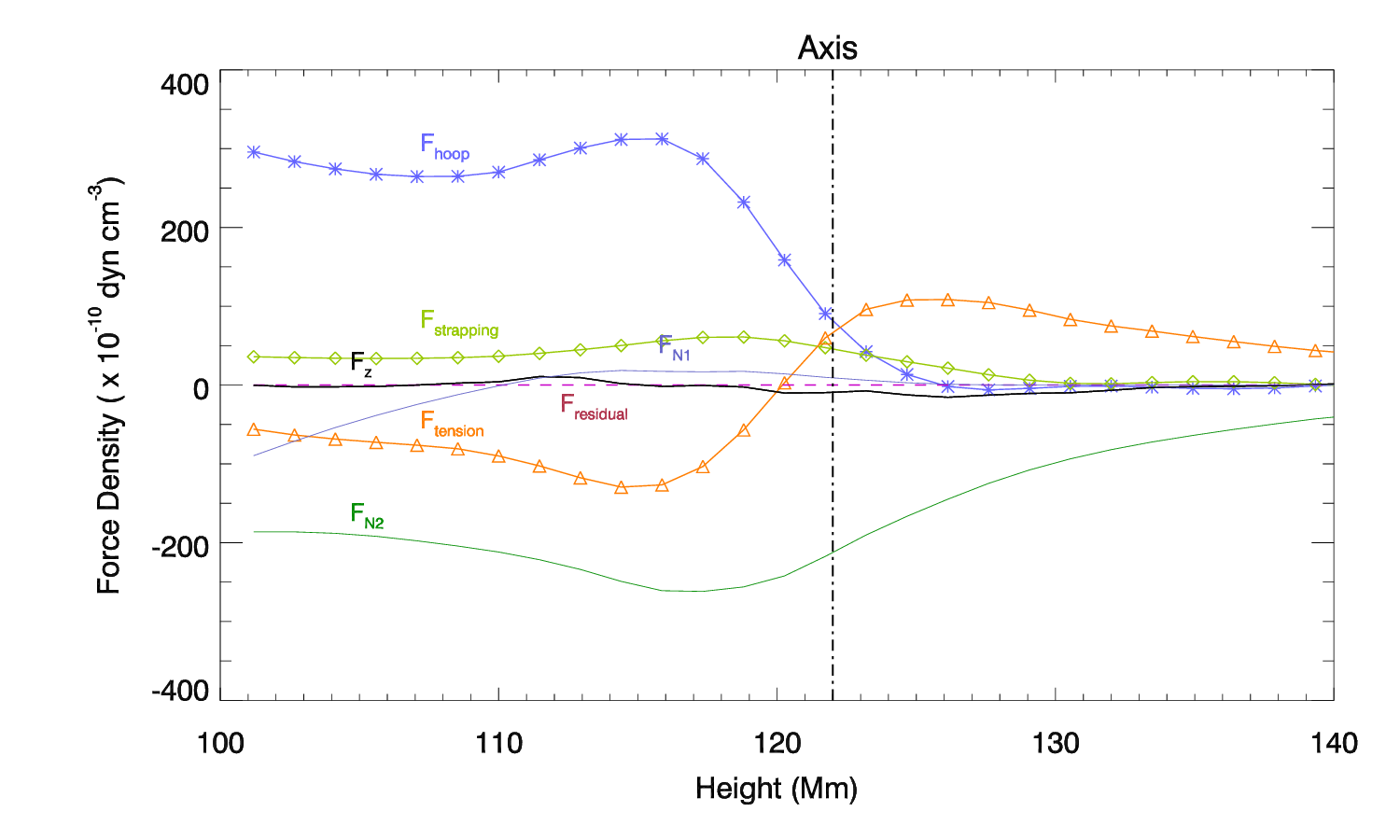}
\caption{Distribution of the total Lorentz force and its components in $z$-dircetion along a vertical line through the apex of MFR1 at 17:48 UT. The thick black solid line shows the total Lorentz force $F_z$. The blue thick line with asterisks, green thick line with diamonds and orange thick line with triangles show the distribution of component forces including the hoop force ($F_{\rm hoop}$), the strapping force ($F_{\rm strapping}$) and the toroidal field tension force ($F_{\rm tension}$), respectively. The bluish-grey (dark green) line shows the distribution of  the non-axisymmetry force $F_{N1}$ ($F_{N2}$). The purple dashed line indicates the residual force ($F_{\rm residual}$) by substracting the other forces from the total Lorentz force $F_z$, which is used to testify the validity of force calculation. The vertical pecked line shows the height of MFR1's axis. }
\label{fig7}
\end{figure}

To understand the trigger mechanism, we check the evolution of the twist in the flux rope as shown Figure \ref{fig6}. Although the twist of MFR1 at the initial time is 1.2, which is larger than the classical value of KI (the Kruskal–Shafranov critical value is N=1), the minimum twist for onset of KI is 1.25 for a line-tied flux rope with a nonuniform twist embedded in an ambient field in our simulation \citep{Kruskal1954,Hood&Priest1981}. Therefore, MFR1 with $|T_w|=1.2$ may not be triggered by KI. Then, we calculate the decay index $n$ to check the role of TI. Compared to Figure 2d in \citet{Zhou2019}, the decay index shows a similar distribution. At around 42 Mm above the photosphere, the decay index reaches the critical value of TI, namely $n=1.5$ \citep{Kliem&Torok2006}. Since the initial flux ropes cover a height range of 15 to 60 Mm, nearly half of the flux rope reaches the height of TI. This fact suggests that TI probably initiates the eruption.

One of the major features of this event is the strong rotation of the filament, which is reproduced fairly well by our simulation. As shown in Figure \ref{fig6}, the mutual transfer between the writhe and twist of the flux rope contains two stages. First, from 17:12 to 17:34 UT, the absolute value of the twist decreases while the writhe increases, corresponding to the rotation process of the eruptive filament. A parameter survey performed by \citet{Kliem2012} reveals that there are two dominant mechanisms for the flux-rope rotation, namely, KI and the Lorentz force produced by the external shear-field component, the latter of which is more liable for the large-angle rotation. In our simulation, the shear-field component provided by the lower flux rope (MFR2) and the external field is the reason for the rotation of the upper flux rope (MFR1). The sheared field interacts with the toroidal current in MFR1 and produces lateral Lorentz forces on two legs of MFR1, which drive the rotation. On the other hand, we find that KI has little effect on the rotation. The initial twist of MFR1 is too small to trigger the eruption. When the twist decreases, the KI is even less likely to work. \citet{Kliem2012} showed that TI could also convert twist to writhe during a flux rope eruption. The conversion is produced by the relaxation of the twist-field magnetic tension, which transforms part of the twist into writhe. 

More intriguingly, similar to \citet{Zhou2022}, we also discover a transfer from the writhe to twist. The difference is that such a transfer occurs in the early phase of the eruption in \citet{Zhou2022}, while it appears in the later period of the confined eruption in our simulation. Moreover, observations show that some failed eruptions of prominences present a reversed rotation about its ascending direction \citep{Song2018}, which might be interpreted by a transfer between writhe and twist. Unfortunately, such a reversal is not captured by observations due to the limitation of spatial-temporal resolution of EUVI, which is 10 minutes for 304 {\AA} passband. However, this reversal can be discovered in our simulation. It's difficult to give an accurate reverse angle because the flux rope becomes incompact during evolution, but there is a reversal around 20 degrees of MFR1 after 17:36 UT. This reversal coinsides with the evolution in Figure \ref{fig6}, where $|W_r|$ decreases and $|T_w|$ increases. Therefore, a conjecture that a similar process occurs in observation is reasonable. For a filament rotation with observations by \emph{SDO}/AIA with 12-second cadence, or EUI on Solar Orbiter with 5-second cadence, there might be a clear sign for a reversed process.

Another interesting characteristic is that both flux rope eruptions are confined. By analyzing the evolution of the Lorentz force, we can understand the dynamics of the flux ropes. First, the total Lorentz force is downward from around 7 Mm below the axis to the upper edge of the flux rope, which leads to a confined eruption. Note that in our simulation, such a force may come from the strong dome-like external flux. By decomposing the Lorentz force, we find that the toroidal field tension force confines the eruption of MFR1's lower half to some extent. More importantly, the non-axisymmetry force $\rm F_{N2}$ plays a crucial role to confine the whole flux rope. 



 We note that, the mechanism of torus instability just considers the strapping force, hoop force, and magnetic tension force generated by the internal toroidal magnetic field, however the toroidal field tension force generated by the external toroidal (guide) field is also significant in determining whether the eruption is successful or not \citep{Myers2015}. For some non-radial eruptions, the rotation of flux ropes may change the poloidal and toroidal components of the external field, which might result in an abrupt increase of the toroidal field tension force and a sign reversal of the strapping force. Considering the deformation of the flux-rope section from the initially circular shape, the force due to the non-axisymmetry, which is an interaction between radial field induced current and the toroidal and poloidal magnetic fields, might also be an important ingredient of the confining force \citep{Zhong2021}. That may be the reason for a close relation between the large-angle rotation and the confined eruption, which has been discovered by observations \citep{Zhou2019}. 

 Here we will provide some explanations on the selection of external fields when calculating Lorentz forces. In the decomposition of Lorentz force, the potential field is used as the external field to compute different force components. This choice is reasonable although it is approximate in some way. The approximations are reflected in the following aspects. First, the normal magnetic field, which is served as the boundary to compute the potential field, is affected by some local electric currents in the corona besides those currents under the photosphere. Second, we neglect the external field of MFR1 contributed by MFR2, which is included as the internal field of MFR1. We didn't separate the magnetic flux contributed by MFR2 because the decompositions considering all electric current distributions in the computation box are really cubersome, and we do not have practical ways to get the results. However, the potential field is an acceptable approximation to the external field especially in the case without filament rotation. In a rotation case, the strapping flux within the erupting rope will rotate with the rope, changing the original strapping flux distribution a lot. Therefore, using the potential field instead of the real external field still has some shortcomings, and hence the individual magnitudes of the hoop and strapping forces, require further study in cases of strong flux rope rotation.

For the failed eruption of MFR2, it is likely to be related to the rotation of MFR1. After a large-angle rotation of MFR1, its toroidal field is almost parallel with the overlaying field, which produces a surge of the strapping field and restricts the rising of MFR2. In most of the previous studies about TI, the decay index is calculated with a potential field extrapolation by assuming quasi-static coronal magnetic fields. However, the flux-rope background field changes dramatically in solar eruptions \citep{Zhong2021}, so the initial decay index cannot forecast the eruption in some complicated cases. Therefore, we attempt to develop a more reasonable dynamic decay index in future works, which is dependent on the flux-rope axis and the background field at different times. We believe that a dynamic decay index might be more compatible with complex events, e.g., rotation, deflection and homologous eruptions. 

Moreover, it is noted that the two adjacent flux ropes have the same sign of the parallel electric current and helicity at the beginning of the eruption, which is prone to trigger the coalescence of two flux ropes \citep{Makwana2018, Linton2001}. However, such an interaction is not evident in our simulation for the following reasons. On the one hand, the flux rope underneath (MFR2) is formed during the eruption, with a minor initial current intensity and faint interaction. On the other hand, the upper flux rope (MFR1) has an upward velocity; therefore, even if there is an attraction between the flux ropes, they may still present a separating motion.

\begin{acknowledgments}

We greatly appreciate the reviewer's suggestions to improve this article. The AIA and HMI data are provided by NASA/\emph{SDO} science teams. The SECCHI data are provided by \emph{STEREO} and the SECCHI consortium. X.M.Z., J.H.G., Y.G., and M.D.D. were supported by the National Key R\&D Program of China (2022YFF0503004, 2021YFA1600504, and 2020YFC2201201) and NSFC (12333009). J.H.G was supported by China Scholarship Council under file No. 202206190140. R.K. was supported by a FWO grant G0B4521N and received funding from the European Research Council (ERC) under the European Union’s Horizon 2020 research and innovation program (grant agreement No. 833251 PROMINENT ERC-ADG 2018) and by Internal funds KU Leuven, project C14/19/089 TRACESpace. The numerical computation was conducted in the High Performance Computing Center (HPCC) at Nanjing University.

\end{acknowledgments}

\bibliography{ms}{}
\bibliographystyle{aasjournal}

\end{document}